\pdfoutput=1
\documentclass{article}

\usepackage{arxiv}

\usepackage[utf8]{inputenc} % allow utf-8 input
\usepackage[T1]{fontenc}    % use 8-bit T1 fonts
\usepackage{hyperref}       % hyperlinks
\usepackage{url}            % simple URL typesetting
\usepackage{booktabs}       % professional-quality tables
\usepackage{amsfonts}       % blackboard math symbols
\usepackage{nicefrac}       % compact symbols for 1/2, etc.
\usepackage{microtype}      % microtypography
\usepackage{graphicx}
\graphicspath{ {./images/} }

\title{Natural Hazards Twitter Dataset}

\author{
 Lingyu Meng \\
  Ingram School of Engineering\\
  Texas State University\\
  San Marcos, TX 78666 \\
  \texttt{l\_m523@txstate.edu} \\
   \And
 Zhijie (Sasha) Dong \\
  Ingram School of Engineering\\
  Texas State University\\
  San Marcos, TX 78666 \\
  \texttt{sasha.dong@txstate.edu} \\
}

\begin{document}
\maketitle
\begin{abstract}
With the development of the Internet, social media has become an important channel for posting disaster-related information. Analyzing attitudes hidden in these texts, known as sentiment analysis, is crucial for the government or relief agencies to improve disaster response efficiency, but it has not received sufficient attention. This paper aims to fill this gap by focusing on investigating attitudes towards disaster response and analyzing targeted relief supplies during disaster response. The contributions of this paper are fourfold. First, we propose several machine learning models for classifying public sentiment concerning disaster-related social media data. Second, we create a natural disaster dataset with sentiment labels, which contains nearly 50,00 Twitter data about different natural disasters in the United States (e.g., a tornado in 2011, a hurricane named Sandy in 2012, a series of floods in 2013, a hurricane named Matthew in 2016, a blizzard in 2016, a hurricane named Harvey in 2017, a hurricane named Michael in 2018, a series of wildfires in 2018, and a hurricane named Dorian in 2019). We are making our dataset available to the research community:\url{https://github.com/Dong-UTIL/Natural-Hazards-Twitter-Dataset}. It is our hope that our contribution will enable the study of sentiment analysis in disaster response. Third, we focus on extracting public attitudes and analyzing the essential needs (e.g., food, housing, transportation, and medical supplies) for the public during disaster response, instead of merely targeting on studying positive or negative attitudes of the public to natural disasters. Fourth, we conduct this research from two different dimensions for a comprehensive understanding of public opinion on disaster response, since disparate hazards caused by different types of natural disasters. 
\end{abstract}

% keywords can be removed
%\keywords{First keyword \and Second keyword \and More}

\section{Introduction}
Big data created from social media like Twitter has made a prominent position in almost all industries and sectors right from individuals to government stakeholders, nongovernment institutions, private businesses, volunteering organizations, and so on. In the recent decade, there has been a spurt of interest in the role of social media data in disaster response, since several natural disasters strike across the globe every year, causing large-scale suffering and economic losses to the public. Although there are many studies about social media and disaster response, respectively, social media data is rarely used for disaster response; thus, experts understand less about information spread via social media in the context of natural hazards and disaster response. In reality, social media data in disaster response can aid in data analytics and information communication by detecting early warning messages, updating the disaster-related data, and monitoring the information sent by the public. Moreover, social media data contains many critical details like the essential relief supplies that the victims lack, the satisfaction that people feel, and the fear that the communities have. Social media data analytics thus makes a new technology that is beneficial for humanitarian organizations and the general public. Importantly, it promises to be an emerging research approach to mitigating the devastation of natural hazards and improving the effectiveness of disaster response.

The purpose of this research is to analyze people's sentimental characteristics during a natural disaster by predicting opinioned texts, as humanitarian organizations can provide targeted assistance based on public attitudes. Since the stimulation of natural disasters will make the public spread their cognitions, opinions, and emotions through the Internet, we can collect bunches of opinioned data. However, it is a formidable task for a human reader to find the sentimental data hidden in a wealth of information, as disaster management is a responsibility for dealing with all humanitarian aspects of emergences that consists of several elements, and people can quickly post related information on social media. Consequently, automated sentiment discovery and summarization systems are needed. Sentiment analysis, sometimes also called opinion mining, grows out of this need. It is a popular subdiscipline of the broader field of Natural Language Processing (NLP), which is concerned with the classification of documents based on the expressed opinions or sentiments of the authors regarding a particular topic. Advantages such as saving capital expenditures (e.g., time, money, and labor), tracking more people satisfactions, and identifying vital sentimental triggers make sentiment analysis become one of the hottest topics for machine learning, especially NLP researchers recently. A common approach for sentiment analysis is proposing machine learning models, by extracting features from textual data and then training the classifier with some known data. However, machine learning models usually are domain specific. In other words, machine learning models do not work well on topics or text genres that are different. Therefore, researchers must develop different machine learning models for their research purposes. As existing research rarely focused on disaster-related studies, this paper aims to fulfill this gap by proposing some machine learning models for natural disasters.

\section{Dataset Description}
\subsection{Natural Disasters Selection}
The natural disasters that are considered in this research include two parts since we intend to study a multidimensional understanding of public opinion on disaster response. We first select five different types of natural disasters that include a Tornado occurred in April 2011, a series of Floods started in September 2013, a Blizzard happened in January 2016, a Hurricane named Harvey in August 2017, and a set of Wildfires burned in August 2018. Secondly, we focus on the identical disaster – hurricane, which is one of the most common and deadliest natural disasters in the United States. To be more specific, they are Hurricane Sandy (October 2012), Hurricane Matthew (September 2016), Hurricane Harvey (August 2017), Hurricane Michael (October 2018), and Hurricane Dorian (August 2019). Details obtained from Wikipedia, like duration of disasters, economic losses, and fatalities, are listed in Table \ref{tab:tab1}.

\begin{table}[h]
 \caption{Details of natural disasters}
  \centering
  \begin{tabular}{cccc}
    \toprule
    Disaster & Duration & Economic Loss (\$ billion) & Fatality\\
    \toprule
    Tornado & 04/25/2011 $\sim$ 04/28/2011 & 11 & 324\\
    Hurricane Sandy & 10/22/2012 $\sim$ 11/02/2012 & 68.7 & 233\\
    Floods & 09/09/2013 $\sim$ 12/31/2013 & 1 & 8\\
    Blizzard & 01/22/2016 $\sim$ 01/24/2016 & 3 & 55\\
    Hurricane Matthew & 09/28/2016 $\sim$ 10/10/2016 & 16.4 & 603\\
    Hurricane Harvey & 08/17/2017 $\sim$ 09/02/2017 & 125 & 107\\
    Wildfires & 08/06/2018 $\sim$ 11/08/2018 & 3.5 & 103\\
    Hurricane Michael & 10/07/2018 $\sim$ 10/16/2018 & 25.1 & 74\\
    Hurricane Dorian & 08/24/2019 $\sim$ 09/10/2019 & 4.68 & 84\\
    \toprule
  \end{tabular}
  \label{tab:tab1}
\end{table}

\subsection{Data Collection}
In this research, Twitter is chosen as the primary sentiment analysis object, as Twitter is a popular microblog that has 140 million active users posting more than 400 million tweets every day. During the period of disaster response, a large number of users posted information like disaster damage reports and disaster preparedness situations, making Twitter an essential social media for updating and accessing data. Mining sentimental data efficiently will better understand the disaster response timely and easily. Twitter has provided an application programming interface (API) that can be used by developers to access and read Twitter data. A streaming API is also offered that can access real-time Twitter data. However, Twitter's search API only allows users to collect 180 requests every 15 minutes in the past seven days, with a maximum number of 100 tweets per claim in the free version. Therefore, this research utilizes TwitterScraper in Python of data collection to retrieve the content and Beautifullsoup4 to parse the retrieved content.

To be more specific, tweets in this paper are collected using keyword filtering techniques, which is a common practice in Twitter analysis in which keywords and dates are two critical inputs for data collection. As this research focuses on extracting public attitudes and analyzing the essential needs (e.g., food, housing, transportation, and medical supplies) for the public during disaster response, we first use different combinations of natural disasters name and essential needs as keywords. Second, to collect more complete data, we extended the tracked time frames by one week before and after the duration of each natural disaster because the government often issues emergency alerts in advance, and the public usually has the lagging pace of information awareness. Also, few related data will be collected when time frames are more than one week. Lastly, we only collected English tweets since it is the major language in the United States. In general, the natural hazards Twitter dataset contains 49,816 tweets. Examples of keywords and time frames used for data collection are list in Table~\ref{tab:tab2}.

\begin{table}[h]
 \caption{Examples of keywords and time frames used for data collection}
  \centering
  \begin{tabular}{ccc}
    \toprule
    Disaster & Keyword & Time Frame\\
    \toprule
    Tornado & Tornado + housing/transportation/food/medical supplies & 04/18/2011 $\sim$ 05/05/2011\\
    Hurricane Sandy & Hurricane Sandy + housing/transportation/food/medical supplies & 10/15/2012 $\sim$ 11/09/2012\\
    Floods & Floods + housing/transportation/food/medical supplies & 09/02/2013 $\sim$ 01/07/2013\\
    Blizzard & Blizzard + housing/transportation/food/medical supplies & 01/15/2016 $\sim$ 01/31/2016\\
    Hurricane Matthew & Hurricane Matthew + housing/transportation/food/medical supplies & 09/21/2016 $\sim$ 10/17/2016\\
    Hurricane Harvey & Hurricane Harvey + housing/transportation/food/medical supplies&  08/10/2017 $\sim$ 09/09/2017\\
    Wildfires & Wildfires + housing/transportation/food/medical supplies & 07/31/2018 $\sim$ 11/15/2018\\
    Hurricane Michael & Hurricane Michael + housing/transportation/food/medical supplies& 10/01/2018 $\sim$ 10/23/2018\\
    Hurricane Dorian & Hurricane Dorian + housing/transportation/food/medical supplies& 08/17/2019 $\sim$ 09/17/2019\\
    \toprule
  \end{tabular}
  \label{tab:tab2}
\end{table}

\subsection{Examples of Sentimental Label}
The determination of positive and negative sentiment will widely differ based on the subject and perspective. To simplify the process, we only study positive and negative attitudes in this paper. Table~\ref{tab:tab3} shows some examples, where positive attitude is assigned the label of 0 and negative attitude is assigned label of 1. 

\begin{table}[h]
 \caption{Examples of sentimental label}
  \centering
  \begin{tabular}{cc}
    \toprule
    Sentimental Label & Tweet\\
    \toprule
    Positive & "I am so happy that the Red Cross offers shelters for us"\\
    Positive & “Uber offers free rides to tornado victims staying in shelters”\\
    Negative & "Stores are empty like no food anywhere"\\
    Negative & “Why the governor has not given any evacuation instructions?”\\
    \toprule
  \end{tabular}
  \label{tab:tab3}
\end{table}

\section{Data \& Access Modalities Introduction}
\subsection{Release v1.0 (April 14, 2020)}
This initial dataset includes two parts. First, we collected data from five different types of natural disasters like a Tornado (April 2011), a series of Floods (September 2013), a Blizzard (January 2016), Hurricane Harvey (August 2017), and a set of Wildfires (August 2018). Second, we collected data from the identical disaster – hurricane, which are Hurricane Sandy (October 2012), Hurricane Matthew (September 2016), Hurricane Harvey (August 2017), Hurricane Michael (October 2018), and Hurricane Dorian (August 2019).

\subsection{Release v1.0 (April 14, 2020)}
The dataset is available at this address: \url{https://github.com/Dong-UTIL/Natural-Hazards-Twitter-Dataset}. The dataset is released in compliance with the Twitter’s Terms \& Conditions. This dataset is still being continuously collected and routinely updated. To request access, interested researchers will need to agree upon the terms of usage dictated by the chosen license. If you have technical questions about the data collection and have any further questions about this dataset , please contact Dr. Zhijie Sasha Dong at sasha.dong@txstate.edu. 

%\bibliography{references}  %%% Remove comment to use the external .bib file (using bibtex).
%%% and comment out the ``thebibliography'' section.

\end{document}